\documentclass[prb,aps,twocolumn,epsfig,floats,showpacs]{revtex4}
\usepackage{graphicx}

\begin{document}

\title{Dynamic correlated Cu(2) magnetic moments in superconducting \\
YBa$_2$(Cu$_{0.96}$Co$_{0.04}$)$_3$O$_y$ ($y \sim$7)}
\author{J.A. Hodges, P. Bonville}
\affiliation{CEA, Centre d'Etudes de Saclay,\\ 
DSM/IRAMIS/Service de Physique de l'Etat Condens\'e, 91191 Gif-sur-Yvette, 
France}
\author{A. Yaouanc, P. {Dalmas de R\'eotier}}
\affiliation{CEA/DSM/Institut Nanosciences et Cryog\'enie, 38054 Grenoble, 
France}
\author{X. Chaud}
\affiliation{CRETA, CNRS, Avenue de Martyrs, 38042 Grenoble, France}
\author{S.P. Cottrell}
\affiliation{ISIS Facility, Rutherford Appleton Laboratory, Chilton, Didcot, 
OX11 0QX, UK} 


\begin{abstract}

We have examined the magnetic properties of superconducting
YBa$_2$(Cu$_{0.96}$Co$_{0.04}$)$_3$O$_y$ ($y\sim$7, T$_{sc}$=65\,K )
using elastic neutron scattering and muon spin relaxation ($\mu$SR) on single
crystal samples. The elastic neutron scattering measurements evidence magnetic
reflections which correspond to a commensurate antiferromagnetic Cu(2) 
magnetic structure with an associated N\'eel temperature T$_N$ $\sim$400\,K. 
This magnetically correlated state is not evidenced by the $\mu$SR 
measurements. 
We suggest this apparent anomaly arises because the magnetically correlated 
state is dynamic in nature. It fluctuates with rates that are low 
enough for it to appear static on the time scale of the elastic neutron 
scattering measurements, whereas on the time scale of the $\mu$SR 
measurements, at least down to $\sim$50\,K, it fluctuates too fast to be 
detected. 
The different results confirm the conclusions reached from work on 
equivalent polycrystalline compounds: the evidenced fluctuating, correlated 
Cu(2) moments coexist at an atomic level with superconductivity.

\end{abstract}


\maketitle

\section {Introduction}

For the cuprates showing high temperature superconductivity, there is much 
interest in how the superconducting properties of the Cu(2)-O planes and 
the Cu(2) based magnetic order evolve as function of the doping level
\cite{orenstein00}. At the present time, the general consensus
is that the ground state involves magnetically ordered Cu(2) for low dopings 
(typically for a doping level, p $<$ $\sim$0.05), 
it involves d-wave superconductivity (for p $\sim$0.05/0.10 to $\sim$0.25) 
and it involves a metallic state for higher dopings. The intermediate regions
notably the underdoped pseudogap regime, where both superconductivity and 
Cu(2) magnetic order may be present, are not yet fully understood. In the 
underdoped region, it is problematic to experimentally determine whether or 
not superconductivity and Cu(2) magnetic order are mutually exclusive at an
atomic level. However, since it is well established that at one end of the 
phase diagram there is region with magnetically ordered Cu(2) and no 
superconductivity and that around optimum doping there is a region with 
superconductivity and no magnetically ordered Cu(2), the two phenomena are
usually considered to be mutually exclusive at an atomic level. Previously 
however \cite{vaast97,hodges02,hodgesa09,isis98}, we have shown that this 
view is not always valid. 

In Ref.\onlinecite{hodgesa09} (preceding paper), we reported that when 
Ni is substituted into fully oxidised YBa$_2$Cu$_3$O$_y$ (y $\sim$7), 
the Cu(2) in the neighbourhood of each Ni carry magnetic moments which are
short range magnetically correlated. For the examined Ni concentration of 4\%,
essentially all the Cu(2) of the sample were found to carry correlated 
magnetic moments. These moments undergo temperature dependent fluctuations 
with rates up to the GHz range.

We have also observed that when Co (rather than Ni) is substituted into 
fully oxidised YBa$_2$Cu$_3$O$_y$ (y $\sim$7), correlated magnetic Cu(2) 
moments are again clearly visible. These results related to elastic neutron
scattering measurements on a single crystal of YBa$_2$Cu$_3$O$_y$ (y $\sim$7) 
substituted with 1.3\% of Co \cite{hodges02} and $^{170}$Yb M\"ossbauer 
\cite{vaast97} and $\mu$SR \cite{isis98} probe measurements on polycrystalline
samples of YBa$_2$Cu$_3$O$_y$ (y $\sim$7) substituted with from 1.0 to 4.0\% 
of Co.
Here, we report additional results, obtained from elastic neutron scattering 
and $\mu$SR measurements, on superconducting single crystal samples of 
YBa$_2$Cu$_3$O$_y$ (y $\sim$7) substituted with 4.0\% Co.  

\section {Samples, Methodology}

The single crystal samples of YBa$_2$(Cu$_{0.96}$Co$_{0.04}$)$_3$O$_y$ 
(y $\sim$ 7) were prepared by the top seeding, melt texturing method and they
were subsequently annealed in oxygen. Because of the growth technique used, 
the samples necessarily contain Y$_2$BaCuO$_5$ as a second phase 
($\sim$40$\%$ by weight) as well as small amounts of some other impurities. 
T$_{sc}$, the superconducting transition temperature, is 65\,K which is 
typical for a 4\% Co substitution level.
The elastic neutron scattering measurements were carried out at the 
Laboratoire L\'eon Brillouin, Saclay, France and
the muon probe measurements were made at the ISIS facility of the 
Rutherford-Appleton Laboratory, Chilton, UK.

The Co enters only the Cu(1) or chain sites and it pulls in additional oxygen
atoms \cite{tarascon88,bridges89,renevier94}. The Co tend to form
dimers \cite{tarascon88,renevier94} or even short chains along the (110) 
direction \cite{bridges89} which introduce microtwinning and trigger the 
change to the tetragonal phase that occurs for Co levels $\sim$0.025. 
The introduction of Co reduces the carrier density 
\cite{clayhold89a,clayhold89b} and the resulting underdoped samples evidence
a pseudo-gap which is seen, for example, in local spin susceptibility 
measurements \cite{dupree92}.
Nuclear quadrupole resonance measurements have shown that the substitution of 
Co into YBa$_2$Cu$_3$O$_7$ introduces magnetic moments at both the chain, 
Cu(1) and plane, Cu(2) sites \cite{matsumura94}. 

\section {Experimental results}

Before presenting the results for the single crystal samples, we briefly
recall the results that have been obtained on the corresponding 
polycrystalline samples.
Both $^{170}$Yb M\"ossbauer \cite{vaast97} and $\mu$SR \cite{isis98}
local probe measurements on superconducting samples of YBa$_2$Cu$_3$O$_y$ 
(y $\sim$7) substituted with Co have shown that the samples contain 
correlated magnetic Cu(2) moments. These moments are introduced over a 
distance of approximately 3 to 4 a/b lattice spacings around the Co.
This length scale is similar to that of the staggered Cu(2) moments 
which are introduced around Ni atoms substituted in YBa$_2$Cu$_3$O$_7$ 
as obtained from NMR measurements \cite{bobroff97,ouazi06}.
The Cu(2) magnetic correlations in superconducting YBa$_2$Cu$_3$O$_y$ 
(y $\sim$7) substituted with Co are short range since the magnetic reflections
are too broad to be seen by neutron diffraction measurements \cite{maury04}. 
The correlated Cu(2) moments are dynamic with temperature dependent rates. The
rates increase as the temperature increases and they extend up to the GHz 
range. At each particular temperature, the local fluctuation rates show a 
distribution. 

These characteristics are quite similar to those evidenced by 
polycrystalline superconducting YBa$_2$Cu$_3$O$_y$ (y $\sim$7) substituted 
with Ni \cite{hodgesa09}, with however one difference: at a given temperature,
the average fluctuation rate of the correlated Cu(2) moments in samples 
containing Co \cite{vaast97,isis98} is considerably lower than that in samples
containing an equivalent amount of Ni \cite{hodgesa09}.

\subsection {Elastic neutron scattering measurements}
\label{mononeut}

The details concerning the measurement procedure are given in ref. 
\onlinecite{hodges02} which reported well defined magnetic reflections in 
superconducting, single crystal YBa$_2$Cu$_3$O$_y$ (y $\sim$7) substituted 
with 1.3\% Co. The reflections corresponded to commensurate
antiferromagnetically correlated Cu(2) magnetic moments with correlation 
lengths typically $>$ $\sim$ 20\,nm. For the superconducting single crystal of
YBa$_2$Cu$_3$O$_y$ (y $\sim$7) substituted with 4\% Co considered here, 
similar well defined magnetic reflections are observed and they again evidence
commensurate antiferromagnetically correlated Cu(2) magnetic moments.
As is discussed below, this magnetically correlated state possesses some, but 
not all, of the characteristics generally associated with conventional 
(static) long range magnetic order. To distinguish the presently reported 
state from that of conventional long range magnetic order, we refer to it as 
an extended range, magnetically correlated state. 

\begin{figure}
\includegraphics[width=0.4\textwidth]{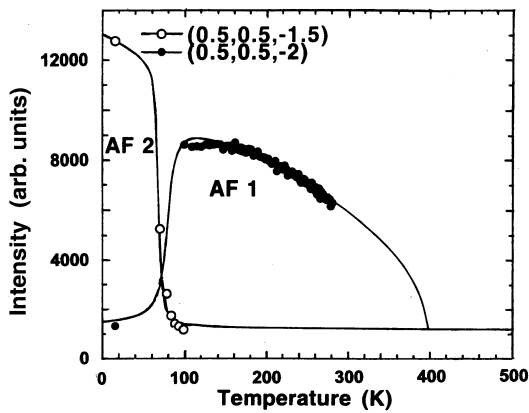}
\caption{From elastic neutron scattering measurements on single crystal
superconducting YBa$_2$(Cu$_{0.96}$Co$_{0.04}$)$_3$O$_7$ (y $\sim$ 7) mixed 
with Y$_2$BaCuO$_5$: 
temperature dependence of the scattering intensity at {\bf Q} = (0.5,0.5,-2) 
which displays a re-entrant behaviour below $\sim$90\,K and at 
{\bf Q} = (0.5,0.5,-1.5) where the intensity grows below $\sim$90\,K. 
The N\'eel temperature, T$_N$ $\sim$400\,K, is obtained by extrapolation as 
described in the text.} 
\label{figconeutmono}
\end{figure}

The temperature dependence of the neutron scattering intensities at  
{\bf Q} = (0.5,0.5,-1.5) and (0.5,0.5,-2) are shown in 
Fig. \ref{figconeutmono}. As the temperature is reduced
below $\sim$300\,K, the intensity at {\bf Q} = (0.5,0.5,-2) initially
increases progressively and then below $\sim$90\,K it decreases. 
Near 90\,K, an intensity appears at (0.5,0.5,-1.5) which 
progressively increases as the temperature is lowered. 
This behaviour indicates that near 90\,K, the system undergoes a
AF1 - AFII transition characterised by a doubling of the AF unit cell along 
the c-axis. A similar doubling of the AF unit cell also occurred in the 
superconducting sample containing 1.3$\%$Co but at a lower temperature 
($\sim$12\,K) \cite{hodges02}. It also occurs in
non-superconducting YBa$_2$Cu$_3$O$_{6+\delta}$ when substituted at the 
Cu(1) site \cite{takatsuka90,brecht95}. In the non-superconducting samples, 
the temperature of the AF1 - AFII transition also increases as the 
concentration of the substituted cation increases.

Although the measurements were made at room temperature and below, this range 
is sufficiently wide to provide a reasonable estimate of the N\'eel 
temperature (T$_N$) which lies above room temperature. The extrapolation was 
made assuming that the thermal dependence of the scattering intensity, 
normalised to T$_N$, is the same as in the sample with 1.3$\%$Co 
\cite{hodges02}. 
The obtained value, T$_N$ $\sim$400\,K, is higher than that for
the sample with 1.3$\%$Co (T$_N$ $\sim$330\,K) \cite{hodges02}.
The maximum size of the correlated Cu(2) magnetic moment 
($\sim$0.3\,$\mu_B$), is approximately twice that ($\sim$0.14\,$\mu_B$) 
observed in the part of the sample containing 1.3$\%$ Co where the moments
are extended range correlated \cite{hodges02}. 

Using elastic neutron scattering, a total of six separately prepared 
superconducting samples of YBa$_2$(Cu$_{1-x}$Co$_x$)$_3$O$_y$ (y $\sim$7) 
have been examined. Each of them evidences well defined magnetic reflections 
corresponding to antiferromagnetically correlated Cu(2) magnetic moments. 
This characteristic is thus very robust. However, in samples having the same 
concentration of Co, the size of the Cu(2) magnetic moment obtained 
assuming all the Cu(2) in the sample contribute to the magnetic 
reflections, is found to vary from sample to sample. 
It was generally lower (by up to 40$\%$) than the value $\sim$0.3\,$\mu_B$ 
reported for the sample above. 

The origin of this sample dependent behaviour may be considered in terms of 
two limiting cases. It could be due to a variation in the size of the extended
range correlated moments (assuming all the Cu(2) contribute to the magnetic 
reflections in each sample) or it could be due to a variation in the volume 
fraction over which the Cu(2) are extended range correlated. In this context
we recall that our local probe measurements on a number of separately
prepared single phase, polycrystalline samples of YBa$_2$Cu$_3$O$_y$ 
(y $\sim$7) substituted with 4\% Co show that in each sample essentially 
all the Cu(2) carry (short range) correlated magnetic moments \cite{vaast97} 
and that the average size of the moment remains essentially the same. 
We thus anticipate that in the single crystal samples, all the Cu(2) also 
carry correlated magnetic moments having a common average size. 
We then attribute the variation of the neutron measured 
Cu(2) magnetic moment from sample to sample to variations in
the relative volumes where the correlations are respectively extended range 
(which contribute to the narrow magnetic reflections) and where they
are short range (which do not contribute to the narrow magnetic reflections).

With this description in terms of different relative volumes of the
extended range correlated and short range correlated fractions,
it also follows that it is possible that not all the 
Cu(2) are extended range correlated in any of the single crystals samples 
including that reported above (Fig. \ref{figconeutmono}). In this case,
the average size of the Cu(2) moment for this sample would be higher 
than the value of $\sim$0.3\,$\mu_B$ obtained assuming that all the Cu(2) 
in the sample are extended range correlated.

We do not know why the correlations are sufficiently extended range in the 
single crystals to give rise to magnetic reflections whereas in the equivalent
polycrystalline samples they remain short range. It is possible 
that the $\sim$40$\%$ of Y$_2$BaCuO$_5$ which is distributed throughout the 
single crystal sample volume could play a role. To test this, it would be of 
interest to examine if the extended range correlations are present in 
equivalent single crystal samples which do not contain any secondary phases. 

The AFI and AFII type magnetic structures we observe in the superconducting 
samples are also present in non-superconducting YBa$_2$Cu$_3$O$_6$ (AFI) 
\cite{burlet88} and in non-superconducting YBa$_2$Cu$_3$O$_6$ substituted at 
the Cu(1) sites (AFI and AFII) \cite{takatsuka90,brecht95}.
Comparing the results obtained here with those in non-superconducting 
YBa$_2$Cu$_3$O$_6$, we note that the N\'eel temperature in the 
superconducting sample ($\sim$400\,K) is only marginally lower than that of 
the non-superconducting sample ($\sim$420\,K) whereas the sample averaged 
size of the Cu(2) magnetic moment in the superconducting sample 
($\sim$0.3\,$\mu_B$) amounts to a considerable fraction of that in 
non-superconducting sample (0.61\,$\mu_B$). 

There are thus a number of similarities between the extended range 
magnetically correlated state observed in superconducting 
YBa$_2$(Cu$_{0.96}$Co$_{0.04}$)$_3$O$_y$ (y $\sim$7) and the conventional
long range magnetically ordered state observed in non-superconducting 
YBa$_2$Cu$_3$O$_6$.
There is however, one important difference. Whereas for 
YBa$_2$Cu$_3$O$_6$, the conventional (static) magnetic order observed by 
neutron measurements \cite{burlet88} is also strongly evidenced by $\mu$SR 
measurements \cite{nishida88} (in fact, the $\mu$SR measurements preceded the
neutron measurements in identifying the existence of magnetic order), for  
single crystal YBa$_2$(Cu$_{0.96}$Co$_{0.04}$)$_3$O$_y$ (y $\sim$7), as is 
described in the following section, the extended range correlated state (both
AFI and AFII structures) evidenced by the neutron scattering measurements is 
not evidenced by the $\mu$SR measurements.

\subsection {$\mu$SR measurements}

The $\mu$SR  measurements were made on a sample where the value of the 
extended range correlated Cu(2) magnetic moment measured by neutron scattering
was $\sim$0.2$\,\mu_B$. 
Since the sample contains $\sim$40\% of Y$_2$BaCuO$_5$ which 
orders magnetically near 15\,K and since the muons are implanted over the 
whole sample volume, the $\mu$SR response will involve contributions from the 
two main constituents as well as from the muons which contribute to the 
background. 
It turns out that it is not possible to unambiguously define the 
contributions coming from each of the two main phases in the sample or to
precisely define the contribution coming from the background.
Because of this, it is not feasible 
to make a detailed quantitative analysis and this is especially the case at 
temperatures near and below the magnetic ordering temperature of 
Y$_2$BaCuO$_5$. Nevertheless, even limiting the analysis to the temperatures 
of 30\,K and above, it is possible to obtain pertinent information.

\begin{figure}
\includegraphics[width=0.4\textwidth]{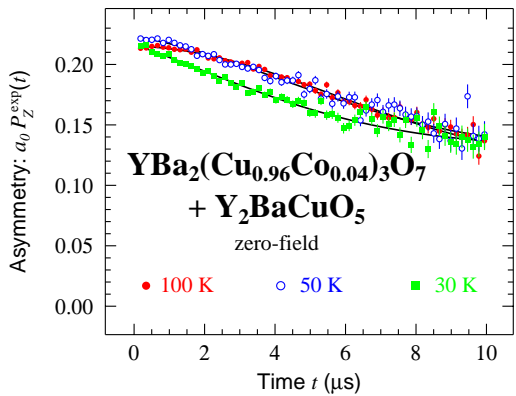}
\caption{(color online) $\mu$SR spectra for single crystal
YBa$_2$(Cu$_{0.96}$Co$_{0.04}$)$_3$O$_y$ (y $\sim$7) mixed with
of Y$_2$BaCuO$_5$ at temperatures above the magnetic ordering temperature
of Y$_2$BaCuO$_5$. 
Despite the fact that neutron scattering measurements on this sample show the 
Cu(2) in YBa$_2$(Cu$_{0.96}$Co$_{0.04}$)$_3$O$_7$ give rise to magnetic 
reflections which correspond to a commensurate antiferromagnetic structure
with an associated N\'eel temperature above room temperature, the 
presence of this correlated state does not markedly influence the 
depolarisation (there is no reduction in the initial asymmetry and there are 
no oscillations).}
\label{figcomuonmono}
\end{figure}

Fig.\ref{figcomuonmono} shows the $\mu$SR depolarisation at 100, 50 and 30\,K
with the incident muons propagating parallel to the c-axis of the single 
crystal. We first describe the simplified approach, in terms of sample 
averaged parameters, which was used to obtain the data fits shown. 
At 100\,K, the time dependent part of total 
depolarisation has the standard Kubo-Toyabe form (a$_s$P$_Z^{KT}$) which is 
characteristic of an interaction between the muon spins and adjacent nuclear 
magnetic moments. The fitted value for the (sample averaged) field width at 
the muon site is $\Delta^{KT}$ = 0.147(3)\,mT. On reducing the temperature to 
50 and 30\,K, the depolarisation shows modest changes.
At 50 and 30\,K, the time dependent part of the depolarisation is well 
described  with a relaxation function 
a$_s$P$_Z^{KT}$exp$({-\lambda_Z t)}$ where $\lambda_Z$ is the (sample 
averaged) muon relaxation rate. Assuming $\Delta^{KT}$ is independent of 
temperature, we obtain (sample averaged) values for $\lambda_Z$ of 
0.051(7)\,$\times$\,10$^6$ (50\,K) and 
0.124(10)\,$\times$\,10$^6$ (30\,K)\,s$^{-1}$. 

We return to these values for $\lambda_z$ below. Here, we consider the
qualitative aspects of the $\mu$SR depolarisation. Down to $\sim$100\,K, the 
depolarisation only evidences interactions between the muon spin and adjacent
nuclear moments and there is no evidence of any interaction between the muon
spin and electron based magnetic moments. An interaction with electron 
based moments is detected near 50\,K and below through the appearance of a 
measurable muon spin relaxation rate whose value increases as the temperature 
is further reduced.
These results pertain to a sample where the major part
(YBa$_2$(Cu$_{0.96}$Co$_{0.04}$)$_3$O$_y$, y $\sim$7) evidences extended range
Cu(2) magnetic correlations with a T$_N$ above room temperature.
The magnetically correlated Cu(2) thus have no (at 100\,K and above) or little
(at 50\,K and below) influence on the muon depolarisation.
We also note that 
whereas conventional (static) long range magnetic order systematically has a 
very marked influence on the $\mu$SR depolarisation (reduction in the initial 
asymmetry or/and oscillations in the time dependence), 
Fig. \ref{figcomuonmono} shows that at each of the three temperatures, the 
initial asymmetry remains at its full value and there are no oscillations. 

We suggest that the neutron scattering evidenced extended range magnetically 
correlated state has little or no influence on the $\mu$SR measurements 
because the correlated state is dynamic: it fluctuates with rates that are 
fast enough to ``motionally narrow'' its influence on the $\mu$SR 
measurements. 

Combined elastic neutron scattering and $\mu$SR measurements on
single crystal YBa$_2$Cu$_3$O$_{6.5}$ (T$_{sc}$ = 55\,K), prepared by the 
top seeding method, have previously identified similar behaviours where
a commensurate Cu(2) antiferromagnetic order with T$_N$ = 310\,K, and 
magnetic moments of  $\sim$ 0.05 $\mu_B$ (assuming a homogeneous distribution
of the moments at all the Cu(2) sites) involved dynamic features with a time 
scale which is longer than that associated with the neutron scattering 
measurements but shorter than that associated with the $\mu$SR measurements
\cite{sidis01}.
The arguments presented there to outline why the moments of 
$\sim$ 0.05 $\mu_B$ would have been readily detected by $\mu$SR if they had 
been ``static'' are also relevant to the present case where the moments 
are approximately four times bigger.

We now return to the (sample averaged) muon spin relaxation rates given above 
since their thermal dependence provides some information concerning the 
dynamically correlated state.
As Y$_2$BaCuO$_5$ magnetically orders near 15\,K, we anticipate that in the
range 30 to 50\,K, it will have paramagnetic behaviour and the local moments
will undergo spin-spin driven fluctuations with rates that are essentially 
independent of temperature. We thus attribute the changes observed in the 
(sample averaged) muon relaxation rate in this temperature range to changes 
that occur in the dynamics of the extended range correlated moments in the 
YBa$_2$(Cu$_{0.96}$Co$_{0.04}$)$_3$O$_y$ (y $\sim$7) fraction. Thus, it is the
slowing down of the fluctuation rate of the correlated Cu(2) moments in the 
superconducting fraction which gives rise to the appearance (near 50\,K) 
and to the increase (below 50\,K) of the muon spin relaxation rate. 
A quite similar appearance (near 50\,K) and a quite similar increase 
(below 50\,K) of the muon spin relaxation rate are also observed in the 
equivalent single phase superconducting polycrystalline samples \cite{isis98} 
where they are clearly related to the properties of the (short range) 
magnetically correlated Cu(2) \cite{vaast97,isis98}. 

\section {Comments}

Although it is not commonplace to encounter a state involving fluctuating 
extended range correlated magnetic moments, in addition to the case of
YBa$_2$Cu$_3$O$_{6.5}$ mentioned above \cite{sidis01}, this state has been 
observed recently in the rare earth pyrochlores where frustration plays a 
role \cite{bertin02,bonvill04,mirebeau05,lago05,wills06,dalmas06,chapuis07}. 
In a manner analogous to that reported here, ref.\onlinecite{dalmas06}
reported both the observation of neutron scattering peaks and of 
``motional narrowing'' of the $\mu$SR response in a particular pyrochlore.
The frequency scale of the fluctuations was of order 
10$^{10}$s$^{-1}$ and the maximum magnetic correlation length 
was $\sim$20nm\cite{mirebeau05}. 
For single crystal YBa$_2$(Cu$_{0.96}$Co$_{0.04}$)$_3$O$_y$ (y $\sim$7), it is
difficult to obtain good estimates of the fluctuation rates which are
probably both strongly temperature dependent and distributed in size. We 
anticipate that depending on the temperature, they could lie in the range from
$\sim$10$^{7}$s$^{-1}$ (the lowest value measured in polycrystalline 
YBa$_2$(Cu$_{0.96}$Ni$_{0.04}$)$_3$O$_y$, y $\sim$7\cite{hodgesa09})
to $\sim$10$^{11}$s$^{-1}$ (observed in polycrystalline 
YBa$_2$(Cu$_{0.96}$Co$_{0.04}$)$_3$O$_y$, y $\sim$7)\cite{vaast97}.

Clearly more experimental work on the single crystals is required (including
samples not containing impurity phases) so as to better define the 
dynamical properties of the correlated Cu(2) moments. The present study simply
suggests that the extended range, commensurate, antiferromagnetic 
Cu(2) state evidenced by the elastic neutron scattering measurements on
superconducting YBa$_2$(Cu$_{0.96}$Co$_{0.04}$)$_3$O$_y$, y $\sim$7 is 
not conventional in that it possesses dynamic characteristics. 

Finally, we note that although the details of the properties of the correlated
Cu(2) magnetic moments are of interest, in the context of the present work, a 
more important aspect is simply that these fluctuating moments exist.

\section{Summary and Conclusions}

When Co is substituted (at the Cu(1) sites) into fully oxidised 
YBa$_2$Cu$_3$O$_y$ (y $\sim$7), the Cu(2) in its neighbourhood 
carry antiferromagnetically correlated moments. 
From measurements on polycrystalline samples,
we previously found a Co substition level of $\sim$2$\%$ was sufficient for
essentially all the Cu(2) to carry magnetic moments (in such samples 
T$_{sc}$ = 84\,K). This indicates that the range around the Co with correlated
Cu(2) moments is approximately three to four a or b lattice spacings.

A mechanism has been proposed to explain the magnetically correlated Cu(2) 
observed in {\it underdoped} cuprates \cite{kilian99}, so that this 
mechanism could play a role here. It is possible however, that a different 
and so far undefined mechanism, is reponsible for the correlated Cu(2) moments
observed both in underdoped (YBa$_2$Cu$_3$O$_7$ + Co) cuprates and in 
optimally doped (YBa$_2$Cu$_3$O$_7$ + Ni) cuprates \cite{hodgesa09}.  

For reasons that remain to be established, the omnipresent fluctuating, 
short range magnetically correlated Cu(2) state observed in polycrystalline  
YBa$_2$(Cu$_{0.96}$Co$_{0.04}$)$_3$O$_y$ (y $\sim$7) becomes extended range in
single crystal samples. 
There are few existing examples of a state involving fluctuating, extended 
range correlated moments but this has been evidenced recently both in 
YBa$_2$Cu$_3$O$_{6.5}$ \cite{sidis01} and in some of the 
rare earth pyrochlores where magnetic frustration plays a key role
\cite{bertin02,bonvill04,mirebeau05,lago05,wills06,dalmas06,chapuis07}.
Whether frustration plays a role in fashioning some of the exotic properties 
of the superconducting cuprates remains to be seen.

The results presented here provide further evidence that the Cu(2)-O network 
of the planes is capable of supporting superconductivity when all the Cu(2) 
carry fluctuating correlated magnetic moments.

\section{Acknowledgements}

J.A.H thanks Philippe Bourges and Yvan Sidis for the neutron scattering 
measurements.

\end{document}